\documentclass[11pt,superscriptaddress,aps,prd,preprint]{revtex4}
\usepackage{amsmath}

\makeatletter
\usepackage[T1]{fontenc}
\usepackage{amsmath}
\usepackage{amssymb}
\usepackage{graphicx}
\usepackage{tabularx}

\usepackage{slashed}

\usepackage{xcolor}

\newcommand{\bea}{\begin{eqnarray}}
\newcommand{\eea}{\end{eqnarray}}

\begin{document}

\title{Spin-1/2 particles in Phase space: Casimir effect and Stefan-Boltzmann law at finite temperature}

\author{R. G. G. Amorim}\email[]{ronniamorim@gmail.com}
\affiliation{Faculdade Gama, Universidade de Bras\'ilia, 70910-900, Bras\'ilia, DF, Brazil}

\author{S. C. Ulhoa}\email[]{sc.ulhoa@gmail.com}
\affiliation{International Center of Physics, Instituto de F\'isica, Universidade de Bras\'ilia, 70910-900, Bras\'ilia, DF, Brazil} \affiliation{Canadian Quantum Research Center, 204-3002, 32 Ave, Vernon, BC V1T 2L7,  Canada.}

\author{J. S da Cruz Filho}\email[]{ze_filho@fisica.ufmt.br}
\affiliation{Secretaria de Estado de Educa\c{c}\~ao de Mato Grosso, 78049-909, Cuiab\'{a}, Mato Grosso, Brazil}

\author{A. F. Santos}\email[]{alesandroferreira@fisica.ufmt.br}
\affiliation{Instituto de F\'{\i}sica, Universidade Federal de Mato Grosso,\\
78060-900, Cuiab\'{a}, Mato Grosso, Brazil}

\author{F. C. Khanna \footnote{Professor Emeritus - Physics Department, Theoretical Physics Institute, University of Alberta\\
Edmonton, Alberta, Canada}}
\email[]{khannaf@uvic.ca}
\affiliation{Department of Physics and Astronomy, University of
Victoria,BC V8P 5C2, Canada}

\begin{abstract}

The Dirac field, spin 1/2 particles, is investigated in phase space. The Dirac propagator is defined. The Thermo Field Dynamics (TFD) formalism is used to introduce finite temperature. The energy-momentum tensor is calculated at finite temperature. The Stefan-Boltzmann law is established and the Casimir effect is calculated for the Dirac field in phase space at zero and finite temperature. A comparative analysis with these results in standard quantum mechanics space is realized.

\end{abstract}

\maketitle

\section{Introduction}

The Wigner function formalism \cite{Wigner,Hillery} and  Noncommutative Geometry \cite{Weyl} play a fundamental role in
the study of phase space quantum mechanics. The Wigner formalism enables
a quantum operator, $A$, defined in the Hilbert space, ${\cal S}$,
to have an equivalent function of the type $a_w(q,p)$,  in phase space $\Gamma$,
 using the Moyal-product or star-product ($\star $). Such a formalism leads to the classical limit of a quantum theory. In fact quantum mechanics is a noncommutative theory whose representation in phase space is an object of debate. The opposite question, i.e., for a given classical function, what is its quantum counterpart? It is solved by using the Weyl transformation which is formulated independent of the phase space. In fact it can be established within the configuration space of the generalized coordinates. The phase space has a well defined physical meaning. The Hamiltonian function is naturally identified with the energy of the system. Establishing a field theory in phase space sheds light on some obscure points in quantum mechanics. For instance the quantum symmetries are better understood in the symplectic structure of phase space which is similar to the role of Lorentz transformation in the covariant formulation of special relativity. This theoretical framework has to include a finite temperature in order to be suitable for experiments.

The star product has been employed  for different objectives. In particular it has been used for
development of a non-relativistic quantum mechanics formalism in terms of a phase space
using the Galilean symmetry representation \cite{Oliveira}. Thus the Schr\"{o}dinger equation is obtained.
In this case, the wave function $\psi =\psi (q,p)$ is a quasi-probability 
amplitude defined in phase space and the Wigner function is obtained in an alternative way i.e. by using $f_{w}(q,p)=\psi ^{\dag }\star \psi $.
The Dirac equation coupled with the electromagnetic field in phase space  \cite{ron1} and applications  \cite{amorimulhoa} have been obtained.

Our goal is to explore the quasi-probability
amplitude to study the effect of temperature using 
Thermo Field Dynamics (TFD) formalism \cite{Kbook, Umezawa1, Umezawa2, Umezawa22, Santana1, Santana2} in a system for spin-$1/2$ particles.
The principles of this theory are the duplication of the Fock space using the Bogoliubov transformations. The TFD formalism is used to study the Casimir effect, at zero and finite temperature. The scalar field in phase space has been studied \cite{SCF} and some exclusive effects have been found at finite temperature. In addition the Stefan-Boltzmann law  for spin-$1/2$ particles in phase space is described in details.

In section II, the symplectic Dirac field is introduced.
In section III, the Thermo Field Dynamics formalism is presented. In section IV, the Stefan-Boltzmann law  is established and the Casimir effect for the Dirac field is calculated in phase space at zero and finite temperature. In the last section, some concluding remarks are presented.

\section{Spin-$1/2$ Field in Phase Space}

A brief outline for spin-$1/2$ particles in phase space formalism is described. For this purpose, the following star operators in phase space are defined
\begin{equation}\label{o1}
\widehat{P}^{\mu}=p^{\mu}\star=p^{\mu}-\frac{i}{2}\frac{\partial}{\partial q_{\mu}},
\end{equation}
 \begin{equation}\label{o2}
\widehat{Q}^{\mu}=q^{\mu}\star=q^{\mu}+\frac{i}{2}\frac{\partial}{\partial p_{\mu}},
\end{equation}
\begin{equation}\label{o3}
\widehat{M}^{\mu\nu}=\widehat{Q}^{\mu}\widehat{P}^{\nu}-\widehat{Q}^{\nu}\widehat{P}^{\mu},
\end{equation}
which satisfy the Heisenberg  commutation relation $[\widehat{Q}^{\mu},\widehat{P}^{\nu}]=ig^{\mu\nu}$, with $g^{\mu\nu}=diag(-1,1,1,1)$. The Poincar\'e algebra has the form
\begin{equation}\label{o4}
 [\widehat{P}^{\mu},\widehat{P}^{\nu}]=0,
 \end{equation}
\begin{equation}\label{o5}
 [\widehat{P}^{\mu},\widehat{M}^{\nu\sigma}]=i(g^{\mu\nu}\widehat{P}^{\sigma}-g^{\sigma\mu}\widehat{P}^{\nu}),
\end{equation}
\begin{equation}\label{o6}
 [\widehat{M}^{\mu\nu},\widehat{M}^{\sigma\rho}]=-i(g^{\mu\rho}\widehat{M}^{\nu\sigma}-g^{\nu\rho}\widehat{M}^{\mu\sigma}+g^{\mu\sigma}\widehat{M}^{\rho\nu}-g^{\nu\sigma}\widehat{M}^{\rho\mu}).
\end{equation}
The operators in Eqs.(\ref{o1}-\ref{o3}) are defined on a Hilbert Space, $\mathcal{H}(\Gamma)$, associated with the phase space $\Gamma$. The operators $\widehat{P}^{\mu}$ and $\widehat{M}^{\mu\nu}$  stand for translations,
rotations and boosts respectively. Functions defined on the Hilbert space  $\mathcal{H}(\Gamma)$ are defined as
\begin{equation}\nonumber
\int \phi^{\ast}(q^{\mu},p^{\mu})\phi(q^{\mu},p^{\mu})dq^{\mu}dp^{\mu}<\infty.
\end{equation}
The Casimir invariants are  $\widehat{P}^{2}=\widehat{P}^{\mu}\widehat{P}_{\mu}$ and  $\widehat{W}=\widehat{W}^{\mu}\widehat{W}_{\mu}$, where $\widehat{W}^{\mu}=1/2\epsilon_{\mu\nu\sigma\rho}\widehat{M}^{\nu\sigma}\widehat{P}^{\rho}$ are Pauli-Lubansky matrices and $\epsilon_{\mu\nu\sigma\rho}$ is the Levi-Civita symbol.

The Dirac equation in phase space is obtained using the invariant operator $\gamma^{\mu}\widehat{P}_{\mu}$. It is defined as
\begin{eqnarray}\label{dirac1}
\gamma^{\mu}\widehat{P}_{\mu}\psi(q,p)&=&m\psi(q,p),\\\nonumber
\gamma^{\mu}\left(p_{\mu}-\frac{i}{2}\frac{\partial}{\partial q^{\mu}}\right)\psi(q,p)&=&m\psi(q,p)\nonumber
\end{eqnarray}
where $\gamma^{\mu}$ are the Dirac matrices. The Lagrangian density for the Dirac equation is
\begin{equation}\label{dirac2}
\mathcal{L}=-\frac{i}{4}\left(\frac{\partial \overline{\psi}}{\partial q^{\mu}}\gamma^{\mu}\psi-\overline{\psi}\gamma^{\mu}\frac{\partial\psi}{\partial q^{\mu}}\right)-\overline{\psi}(m-\gamma^{\mu}p_{\mu})\psi,
\end{equation}
where $\overline{\psi}=\psi^{\dagger}\gamma^{0}$ and $m$ is the mass of the particle.

The Wigner function provides the physical interpretation \cite{ron1} and is given as
\begin{equation}\label{wigner1}
f_{W}(q,p)=\overline{\psi}(q,p)\star\psi(q,p),
\end{equation}
where the star product is defined by
\begin{equation}\nonumber
a_{W}(q,p)\star b_{W}(q,p)=a_{W}(q,p)\exp \left[ \frac{i\hbar }{2}\left(\frac{%
\overleftarrow{\partial }}{\partial q}\frac{\overrightarrow{\partial }}{%
\partial p}-\frac{\overleftarrow{\partial }}{\partial p}\frac{%
\overrightarrow{\partial }}{\partial q}\right)\right] b_{W}(q,p).  \nonumber
\end{equation}%
Using the Noether theorem in phase space \cite{ron1}, the energy-momentum tensor for the Dirac field is
\begin{equation}\label{dirac3}
\theta_{D}^{\mu\nu}=-\frac{i}{4}\left(-\overline{\psi}\gamma^{\mu}\frac{\partial\psi}{\partial q_{\nu}}+\gamma^{\mu}\psi\frac{\partial\overline{\psi}}{\partial q_{\nu}}\right)-g^{\mu\nu}\mathcal{L}.
\end{equation}

Then, the Green function, $G_D(q-q',p-p')$ is defined as
\begin{equation}\label{dirac6}
\frac{i}{2}\gamma^{\mu}\frac{\partial G_D(q-q',p-p')}{\partial q^{\mu}}+(m-\gamma^{\mu}p_{\mu})G_D(q-q',p-p')=\delta(q-q')\delta(p-p')\,,
\end{equation}
 which may be written as
\begin{equation}\label{dirac8}
\frac{1}{2}\gamma^{\mu} k_\mu \widetilde{G}(k,p-p')+(m-\gamma^{\mu}p_{\mu})\widetilde{G}(k,p-p')=\delta(p-p'),
\end{equation}
where $\widetilde{G}(k,p-p')=\frac{1}{(2\pi)^4}\int d^4q e^{ik^{\mu}(q_{\mu}-q'_{\mu})}G_D(q-q',p-p')$. The propagator of the Dirac field is 
\begin{equation}\label{dirac9}
\widetilde{G}(k,p-p')=\frac{\delta(p-p')}{\gamma^{\mu}\left[\frac{k_\mu}{2}-p_{\mu}\right] +m}.
\end{equation}
Then eq. (\ref{dirac8}) is an algebraic expression which yields (\ref{dirac9}). Then the Green function, $G_D$, is 
\bea
G_D(q-q',p-p')&=&\int \frac{d^4k}{(2\pi)^4} e^{-ik^{\mu} (q_{\mu}-q'_{\mu})}\widetilde{G}(k,p-p')\nonumber\\
&=&\int \frac{d^4k}{(2\pi)^4} \frac{e^{-ik^{\mu} (q_{\mu}-q'_{\mu})}\,\delta(p-p')}{\gamma^{\mu}\left[ \frac{k_\mu}{2}-p_{\mu}\right] +m}.
\eea
Taking $M=2m$, the expression is
\bea
G_D(q-q',p-p')&=& 2e^{-2i(q^{\mu}-q'^{\mu})(p_{\mu}-p'_{\mu})}\,\delta(p-p')\int \frac{d^4k}{(2\pi)^4} \frac{e^{-ik^{\mu} (q_{\mu}-q'_{\mu})}}{\gamma^{\mu}k_\mu+M}\,.
\eea
This leads to the Green function for the Dirac equation in phase space
\bea
G_D(q-q',p-p')&=& 2e^{-2i(q^{\mu}-q'^{\mu})(p_{\mu}-p'_{\mu})}\,\delta(p-p')\left(i\partial_\mu\gamma^\mu -M\right)G_0(q-q')\,,\label{greend}
\eea 
$G_0(q-q')$ is defined as
$$G_0(q-q')=-\frac{iM}{4\pi^2}\,\frac{K_1\left(\sqrt{-(q^{\mu}-q'^{\mu})(q_{\mu}-q'_{\mu})}\,\right)}{\sqrt{-(q^{\mu}-q'^{\mu})(q_{\mu}-q'_{\mu})}}\,,$$ where $\kappa_\nu(q)$ is the modified Bessel function. It should be noted that due to the dependence on the Dirac matrices the Green's function has matrix properties itself.

\section{Thermo Field Dynamics formalism}

 The Thermo Field Dynamics (TFD) is a thermal quantum field theory at finite temperature \cite{Kbook, Umezawa1, Umezawa2, Umezawa22, Santana1, Santana2}. It has two basics elements: (i) doubling the degrees of freedom in a Hilbert space and (ii) the Bogoliubov transformation. The doubling of Hilbert space is given by the tilde ($^\thicksim$) conjugate rules where the thermal space is ${\cal S}_T={\cal S}\otimes \tilde{\cal S}$, with ${\cal S}$ being the standard Hilbert space and $\tilde{\cal S}$ the tilde (dual) space. There is a mapping between the two spaces, i.e., the map between the tilde $\tilde{b_i}$ and non-tilde $b_i$ operators is defined by the following tilde conjugation rules:
\bea
(b_ib_j)^\thicksim = \tilde{b_i}\tilde{b_j}, \quad\quad (cb_i+b_j)^\thicksim = c^*\tilde{b_i}+\tilde{b_j}, \quad\quad (b_i^\dagger)^\thicksim = \tilde{b_i}^\dagger, \quad\quad (\tilde{b_i})^\thicksim = -\xi b_i,
\eea
with $\xi = -1$ for bosons and $\xi = +1$ for fermions.

The Bogoliubov transformation corresponds to a rotation of the tilde and non-tilde variables. Using the doublet notation, for fermions leads to
\bea
b^a=\left( \begin{array}{cc} b^1(\alpha)  \\ b^2(\alpha) \end{array} \right)=\left( \begin{array}{cc} b(\alpha)  \\ \tilde b^\dagger(\alpha) \end{array} \right)={\cal B}(\alpha)\left( \begin{array}{cc} b(k)  \\ \tilde b^\dagger(k) \end{array} \right),
\eea
where $(b^\dagger, \tilde{b}^\dagger)$ are creation operators, $(b, \tilde{b})$ are destruction operators and ${\cal B}(\alpha)$ is the Bogoliubov transformation given by
\bea
{\cal B}(\alpha)=\left( \begin{array}{cc} u(\alpha) & -v(\alpha) \\
v(\alpha) & u(\alpha) \end{array} \right).
\eea
Taking $\alpha=\beta$ ($\beta\equiv1/k_B T$ with $k_B$ being the Boltzmann constant and $T$ the temperature) the thermal operators are written explicitly as
\bea
b(\beta)&=&u(\beta)b(k)-v(\beta)\tilde b^\dagger(k)\\
\tilde{b}(\beta)&=&u(\beta)\tilde{b}(k)+v(\beta) b^\dagger(k)\\
b^\dagger(\beta)&=&u(\beta)b^\dagger(k)-v(\beta)\tilde b(k)\\
\tilde{b}^\dagger(\beta)&=&u(\beta)\tilde{b}^\dagger(k)+v(\beta) b(k).
\eea
These thermal operators satisfy the algebraic rules
\begin{eqnarray}\label{2.5}
&&\{ b_{p}(\beta), b^\dagger_{q}(\beta) \}=\delta^3(p-q),\\
&&\{ \tilde{b}_{p}(\beta), \tilde{b}^\dagger_{q}(\beta) \}=\delta^3(p-q),
\end{eqnarray}
and other anti-commutation relations are null. In addition, the quantities $u(\beta)$ and $v(\beta)$ are related to the Fermi distribution, i.e.,
\bea
u^2(\beta)=\frac{1}{1+e^{-\beta \omega}} \quad \mathrm{and} \quad v^2(\beta)=\frac{1}{1+e^{\beta \omega}}
\eea
such that  $v^2(\beta)+u^2(\beta)=1$.  The parameter $\alpha$ is associated with temperature, but, in general, it may be associated with other physical quantities. In general, a field theory on the topology $\Gamma_D^d=(\mathbb{S}^1)^d\times \mathbb{R}^{D-d}$ with $1\leq d \leq D$, is considered. $D$ are the space-time dimensions and $d$ is the number of compactified dimensions. This establishes a formalism such that any set of dimensions of the manifold $\mathbb{R}^{D}$ can be compactified, where the circumference of the $nth$ $\mathbb{S}^1$ is specified by $\alpha_n$. The $\alpha$ parameter is assumed as the compactification parameter defined by $\alpha=(\alpha_0,\alpha_1,\cdots\alpha_{D-1})$. The effect of temperature is described by the choice $\alpha_0\equiv\beta$ and $\alpha_1,\cdots\alpha_{D-1}=0$.

 Any field in the TFD formalism may be written in terms of the $\alpha$-parameter. As an example the scalar field is considered. Then the $\alpha$-dependent scalar field becomes
\bea
\phi(q, p;\alpha)&=&{\cal B}(\alpha)\phi(q, p){\cal B}^{-1}(\alpha),
\eea
where the Bogoliubov transformation is used.

The $\alpha$-dependent propagator for the scalar field is 
\bea
G_0^{(ab)}(q-q', p-p';\alpha)=i\langle 0,\tilde{0}| \tau[\phi^a(q, p;\alpha)\phi^b(q', p';\alpha)]| 0,\tilde{0}\rangle,
\eea
where $\tau$ is the time ordering operator. Using $|0(\alpha)\rangle={\cal B}(\alpha)|0,\tilde{0}\rangle$ leads to the Green function
\bea
G_0^{(ab)}(q-q', p-p';\alpha)&=&i\langle 0(\alpha)| \tau[\phi^a(q, p)\phi^b(q', p')]| 0(\alpha)\rangle,\nonumber\\
&=&i\int \frac{d^4k}{(2\pi)^4}e^{-ik(q-q')(p-p')}G_0^{(ab)}(k;\alpha),
\eea
where
\bea
G_0^{(ab)}(k;\alpha)={\cal B}^{-1}(k;\alpha)G_0^{(ab)}(k){\cal B}(k;\alpha),
\eea
with ${\cal B}(k;\alpha)$ being the Bogoliubov transformation and
\bea
G_0^{(ab)}(k)=\left( \begin{array}{cc} G_0(k) & 0 \\
0 & G^*_0(k) \end{array} \right),
\eea
 where 
\bea
G_0(k)=\frac{1}{k^2-m^2+i\epsilon},
\eea
is the scalar field propagator and $m$  is the scalar field mass. Here $G^*_0(k)$ is the complex conjugate of $G_0(k)$.

It is important to note that, the physical quantities are given by the non-tilde variables. Then the physical Green function $G_0^{(11)}(k;\alpha)$ is written as
\bea
G_0^{(11)}(k;\alpha)&\equiv & G_0(k;\alpha)=G_0(k)+v^2(\alpha)[G^*_0(k) - G_0(k)].
\eea
where
\bea
v^2(k;\alpha)=\sum_{s=1}^d\sum_{\lbrace\sigma_s\rbrace}2^{s-1}\sum_{l_{\sigma_1},...,l_{\sigma_s}=1}^\infty(-\eta)^{s+\sum_{r=1}^sl_{\sigma_r}}\,\exp\left[{-\sum_{j=1}^s\alpha_{\sigma_j} l_{\sigma_j} k^{\sigma_j}}\right],\label{BT}
\eea
is the generalized Bogoliubov transformation \cite{GBT}, with $d$ being the number of compactified dimensions, $\eta=1(-1)$ for fermions (bosons), $\lbrace\sigma_s\rbrace$ denotes the set of all permutations with $s$ elements and $k$ is the 4-momentum. In the next section, three different topologies are used~\cite{berti}. (i) The topology $\Gamma_4^1=\mathbb{S}^1\times\mathbb{R}^{3}$, where $\alpha=(\beta,0,0,0)$. In this case the time-axis is compactified in $\mathbb{S}^1$, with circumference $\beta$. (ii) The topology $\Gamma_4^1$ with $\alpha=(0,0,0,i2d)$, where the compactification along the coordinate $z$ is considered. (iii) The topology $\Gamma_4^2=\mathbb{S}^1\times\mathbb{S}^1\times\mathbb{R}^{2}$ with $\alpha=(\beta,0,0,i2d)$ is used. In this case the double compactification consists in time and the coordinate $z$. Then thermal effects are considered for Casimir effect and Stefan-Boltzmann Law.

\section{Stefan-Boltzmann law and Casimir Effect for the Dirac field in phase space }

The Stefan-Boltzmann law is calculated by analyzing the energy-momentum tensor given as 
\bea
\theta_{D}^{\mu\nu}(q,p)&=&\lim_{(q',p')\rightarrow (q,p)}\mathbf{\tau}\Biggl\{ -\frac{i}{4}\left[-\overline{\psi}'\gamma^{\mu}\frac{\partial\psi}{\partial q_{\nu}}+\gamma^{\mu}\psi\frac{\partial\overline{\psi}'}{\partial q_{\nu}'}\right]\nonumber\\
&+&g^{\mu\nu}\left[\frac{i}{4}\left(\frac{\partial \overline{\psi}}{\partial q^{\lambda}}\gamma^{\lambda}\psi'-\bar{\psi}\gamma^\lambda\frac{\partial \psi'}{\partial {q^\lambda}'} \right)+\overline{\psi}(m-\gamma^{\mu}p_{\mu})\psi'\right]\Biggl\},\nonumber\\
&=&\lim_{(q',p')\rightarrow (q,p)}\left\{\Gamma^{\mu\nu}\tau\left[\overline{\psi}'(q',p')\psi(q,p)\right]\right\},
\eea
where
\bea
\Gamma^{\mu\nu}=-\frac{i}{4}\left[-\gamma^\mu\frac{\partial}{\partial q_\nu}+\gamma^\mu\frac{\partial}{\partial q_\nu'}-g^{\mu\nu}\left(\frac{\partial}{\partial q^\lambda}\gamma^\lambda-\gamma^\lambda\frac{\partial}{\partial {q^\lambda}'} \right) \right] +g^{\mu\nu}(m-\gamma^\mu p_\mu)\,.
\eea
It should be noted that the field $\psi'$ is the dirac field in phase space as a function of the variables (q',p'), i. e., $\psi' \equiv \psi(q',p')$. The vacuum expectation value of the energy-momentum tensor is 
\bea
\langle \mathcal{\theta_D^{\mu \nu}}(q,p)\rangle=\lim_{(q',p')\rightarrow (q,p)}\left\{\Gamma^{\mu\nu}\left\langle 0\Bigl| \tau\left[\overline{\psi}'(q',p')\psi(q,p)\right]\Bigl| 0\right\rangle\right\}.
\eea
The Dirac propagator in phase space is defined in eq. (\ref{greend}) as
\bea
G_D(q-q',p-p')&=&i\left\langle 0\Bigl| \tau\left[\overline{\psi}'(q',p')\psi(q,p)\right]\Bigl| 0\right\rangle\,.
\eea 
Then the energy-momentum tensor has the form
\bea
\left\langle \mathcal{\theta_D^{\mu \nu}}(q,p)\right\rangle=-i\lim_{(q',p')\rightarrow (q,p)}\left\{\Gamma^{\mu\nu}\left[2e^{-2i(q^{\mu}-q'^{\mu})(p_{\mu}-p'_{\mu})}\,\delta(p-p')\right] \left(i\partial_\mu\gamma^\mu -2m\right)G_0(q-q')\right\}.
\eea

The vacuum average of the energy-momentum tensor in terms of $\alpha$-dependent fields becomes
\bea
\left\langle \mathcal{\theta_D^{\mu \nu}}^{(ab)}(q,p;\alpha)\right\rangle=\lim_{(q',p')\rightarrow (q,p)}\left\{-i\Gamma^{\mu\nu}\left[2e^{-2i(q^{\mu}-q'^{\mu})(p_{\mu}-p'_{\mu})}\,\delta(p-p')\right] \left(i\partial_\mu\gamma^\mu -2m\right)G_0(q-q';\alpha)\right\}\,.\nonumber\\
\eea
In order to obtain measurable physical quantities at finite temperature, a renormalisation procedure is carried out. The physical energy-momentum tensor is defined as
\bea
{\cal T}^{\mu\nu (ab)}(q,p;\alpha)&=&\langle \theta_D^{\mu\nu(ab)}(q,p;\alpha)\rangle-\langle \theta_D^{\mu\nu(ab)}(q,p)\rangle\nonumber\\
&=&-i \lim_{(q',p')\rightarrow (q,p)}\left\{\Gamma^{\mu\nu}\overline{G}_D^{(ab)}(q-q', p-p'; \alpha)\right\},
\eea
where
\bea
\overline{G}_D^{(ab)}(q- q', p - p';\alpha)&=&G_D^{(ab)}(q- q', p - p';\alpha)-G_D^{(ab)}(q- q', p - p').
\eea
Now the Stefan-Boltzmann Law and the Casimir effect in phase space are calculated at finite temperature.

\subsection{Stefan-Boltzmann Law}
The study of the Stefan-Boltzmann law in phase space corresponds to a choice of the parameter $\alpha$. It is important to note that the parameter $\alpha$ is the compactification parameter that is defined as $\alpha=(\alpha_0,\alpha_1,\cdots\alpha_{D-1})$. The temperature effect is described by the choice $\alpha=(\beta, 0, 0, 0)$

The generalized Bogoliubov transformation, Eq. (\ref{BT}) for these parameters is
\bea
v^2(k,\beta)=\sum_{l=1}^\infty (-1)^{l+1}\,e^{-\beta k_0 l}\,.
\eea
The Green's function for the Dirac field in phase space is 
\bea
\overline{G}_D^{(ab)}(q- q',  p - p';\beta)&=&\sum_{l=1}^\infty (-1)^{l+1}\bigl[G^*_D(q-q'+i\beta l n_0, p-p')\nonumber\\
&-&G_D(q-q'-i\beta l n_0,p-p')\bigl],
\eea
where $n_0=(1,0,0,0)$ is a time-like vector. Then the physical energy-momentum tensor is 
\bea
{\cal T}^{\mu\nu (11)}(\beta)&=& -i\lim_{(q',p')\rightarrow (q,p)}\sum_{l=1}^\infty (-1)^{l+1}\Gamma^{\mu\nu}\bigl[G^*_D(q-q'+i\beta l n_0, p-p')\nonumber\\
&-&G_D(q-q'-i\beta l n_0,p-p')\bigl].
\eea
In order to calculate the Stefan-Boltzmann law, take $\mu=\nu=0$, leads to
\bea
{\cal T}^{00 (11)}(\beta)&=& -\lim_{p'\rightarrow p}\sum_{l=1}^\infty\frac{4m \delta(p-p')(-1)^{l+1}e^{-2(p_0-p'_0)\beta l}}{\pi^2l^3\beta^3}\Biggl\{m(l\beta)^2( p_0-p'_0)\kappa_0(2m l\beta)\nonumber\\
&+&K_1(2m l\beta)\Biggl[3+2(m l\beta)^2+l\beta (p_0-p'_0)(1-m l\beta\gamma^0)\Biggl]\Biggl\}.
\eea
This is the Stefan-Boltzmann law for the Dirac field in phase space. It is worth  pointing out that the result ${\cal T}^{00 (11)}(\beta)\sim T^4$ is recovered by taking the limit of the momentum variable. This result in phase space is necessary to compare with experiment. In this sense we can integrate over the momenta which explicitly yields

$${\cal T}^{00 (11)}(\beta)= -\sum_{l=1}^\infty\frac{4m (-1)^{l+1}\beta l}{\pi^2l^3\beta^3}\Biggl\{K_1(2m l\beta)\Biggl[3+2(m l\beta)^2\Biggl]\Biggl\}\,,$$
and take the limit $m\rightarrow 0$, then the only remaining part is the factor of $K_1(2m l\beta)$. That leads to the dependency $\beta^{-4}=T^4$, once the limit of Bessel function is taken. On the other hand it is possible to project in the momentum space by integrating over coordinates. This process leads to a divergence which is of the same nature of the coordinate projection in the absence of temperature. Hence it is necessary a quantity in the momentum space analogous to the temperature, that is the thermal energy.  The introduction of TFD formalism introduces the role of temperature, but it can equaly do the same for the thermal energy. Using phase space and TFD allows us to deal with systems where microscopic energy is dominant.

\subsection{Casimir Effect for the Dirac Field in Phase Space}

Here the choice is $\alpha=(0,0,0,i2d)$, then
\bea
v^2(d)=\sum_{l_3=1}^{\infty} (-1)^{l_3+1}\,e^{-i2d k^3l_3}.
\eea
The Green function is this case is
\bea
\overline{G}_D^{(ab)}(q- q',  p - p';d)&=&\sum_{l_3=1}^\infty (-1)^{l_3+1}\bigl[G^*_D(q-q'+2dl_3 n_3, p-p')\nonumber\\
&-&G_D(q-q'-2dl_3 n_3,p-p')\bigl],
\eea
where $n_3=(0,0,0,1)$ is a space-like vector. Then the energy-momentum tensor is
\bea
{\cal T}^{\mu\nu (11)}(d)&=&-i\lim_{(q',p')\rightarrow (q,p)}\sum_{l_3=1}^\infty (-1)^{l_3+1}\Gamma^{\mu\nu}\bigl[G^*_D(q-q'+2dl_3 n_3, p-p')\nonumber\\
&-&G_D(q-q'-2dl_3 n_3,p-p')\bigl].
\eea

By taking $\mu=\nu=0$, the Casimir energy for the Dirac field in phase space at zero temperature is
\bea
{\cal T}^{00 (11)}(d)&=& \lim_{p'\rightarrow p}\sum_{l_3=1}^\infty\frac{m^2 \delta(p-p')(-1)^{l_3+1}\, e^{4i(p_z-p'_z)dl_3}}{\pi^2 dl_3}\left[1+2(p_0-p'_0)\gamma^0\right] K_1(2m dl_3)\,.
\eea
And for $\mu=\nu=3$, the Casimir pressure in phase space is
\bea
{\cal T}^{33 (11)}(d)&=& \lim_{p'\rightarrow p}\sum_{l_3=1}^\infty\frac{m^2 \delta(p-p')(-1)^{l_3+1}\, e^{4i(p_z-p'_z)dl_3}}{\pi^2 dl_3}\Biggl\{\left[im\gamma_3+(p_z-p'_z)\right] \kappa_2(2m dl_3)\nonumber\\
&-& 2\left[m-(p_z-p'_z)\gamma^3\right] K_1(2m dl_3)\Biggr\}\,.
\eea
 It reproduces the usual result when $m\rightarrow 0$ and integrated over the momenta which means the projection on coordinates space. Then only the factors of $K_1$ is left, the limit of this part yields the dependency $d^{-4}$. Here the dependency on $\gamma$ matrices should be viewed as part of the phase space formalism which is by its core matricial. This part doesn't survive once the projection on coordinates is performed but it is part of the behavior in phase space. In order to be compared with experimental data the projection on momenta space requires the introduction the thermal energy. 

\subsection{Casimir Effect for the Dirac Field in Phase Space at finite temperature}

The effect of temperature is introduced by taken $\alpha=(\beta,0,0,i2d)$. Then the generalized Bogoliubov transformation becomes
\bea
v^2(\beta,d)&=&\sum_{l=1}^\infty (-1)^{l+1}e^{-\beta k^0l}+\sum_{l_3=1}^\infty(-1)^{l_3+1} e^{-i2dk^3l_3}+2\sum_{l,l_3=1}^\infty (-1)^{l+l_3}e^{-\beta k^0l-i2dk^3l_3}.\label{BT3}
\eea
The first two terms of these expressions correspond, respectively, to the Stefan-Boltzmann term and the Casimir effect at $T = 0$. The third term is analyzed and it leads to the Green function
\bea
\overline{G}_D^{(ab)}(q- q',  p - p';\beta, d)&=&\sum_{l_3=1}^\infty (-1)^{l+l_3}\bigl[G^*_D(q-q'+i\beta l n_0+2dl_3 n_3, p-p')\nonumber\\
&-&G_D(q-q'-i\beta l n_0-2dl_3 n_3,p-p')\bigl].
\eea
Then the Casimir energy at finite temperature is
{\small
\bea
{\cal T}^{00 (11)}(\beta, d)&=& \lim_{p'\rightarrow p}\sum_{l_0,l_3=1}^\infty\frac{2m \delta(p-p')(-1)^{l_3+l_0}\, e^{-4i(p_z-p'_z)dl_3-2(p_0-p'_0)\beta l_0}}{4\pi^2 \left(4d^2l_3^2+\beta^2l_0^2\right)^2}\Biggr\{\kappa_0\left(m\sqrt{4d^2l_3^2+\beta^2l_0^2}\right)\nonumber\\
&&2m\left[(2dl_3)^2(1+2m\beta l_0)-(\beta l_0)^2(3-2m\beta l_0)-24id\beta l_0l_3\gamma^3\right]+ \frac{K_1\left(m \sqrt{4d^2l_3^2+\beta^2l_0^2}\right)}{\left(4d^2l_3^2+\beta^2l_0^2\right)^{1/2}}\nonumber\\
&&[-4m(2d^2l_3^2)^2-(\beta l_0)^2\left[12-2m\beta l_0(4-2m\beta l_0)\right]+2(dl_3)^2\left[8-2m\beta l_0(-8+6m\beta l_0)\right]\nonumber\\
&-&48im^2d^3l_3^3\beta l_0\gamma^3-3id\beta l_0l_3\gamma^3(32+4m^2\beta^2 l_0^2)]\Biggl\}\,,
\eea}
and the Casimir pressure at finite temperature is

\bea
{\cal T}^{33 (11)}(\beta, d)&=&  \lim_{p'\rightarrow p}\sum_{l_0,l_3=1}^\infty\frac{2m \delta(p-p')(-1)^{l_3+l_0}\, e^{-4i(p_z-p'_z)dl_3-2(p_0-p'_0)\beta l_0}}{4\pi^2 \left(4d^2l_3^2+\beta^2l_0^2\right)^2}\Biggr\{\kappa_0\left(m \sqrt{4d^2l_3^2+\beta^2l_0^2}\right)\nonumber\\
&&4m\left[(\beta l_0)^2+4dl_3(5dl_3+6i\beta l_0\gamma^3)\right]+ \frac{K_1\left(m \sqrt{4d^2l_3^2+\beta^2l_0^2}\right)}{\left(4d^2l_3^2+\beta^2l_0^2\right)^{1/2}}[128m^2(d^2l_3^2)^2\nonumber\\
&+&160 d^2l_3^2+(\beta l_0)^2(8+48m^2 d^2l_3^2)+4m^2(\beta l_0)^4+6id\beta l_0l_3(32+4m^2(4d^2l_3^2+\beta^2l_0^2))\gamma^3]\nonumber\\
&-&4m(4d^2l_3^2+\beta^2l_0^2)[2+2m\beta l_0+2i(4mdl_3+i(p_z-p_z')\beta l_0)\gamma^3]\nonumber\\
&& \kappa_2\left(m \sqrt{4d^2l_3^2+\beta^2l_0^2}\right)\Biggl\}\,.
\eea
It should be noted that in the limit $p\rightarrow p'$ both the Casimir energy and pressure are real quantities at zero and finite temperature. In the limit $\beta\rightarrow 0$, i.e., $T\rightarrow \infty$, the Casimir energy and pressure become
\bea
{\cal T}^{00 (11)}(d)&=& \lim_{p'\rightarrow p}\sum_{l_3=1}^\infty\frac{m \delta(p-p')(-1)^{l_3+l_0}\, e^{-4i(p_z-p'_z)dl_3}}{4\pi^2 \left(dl_3\right)^3}\Biggl[dl_3\kappa_0(2mdl_3)\nonumber\\
&+&\Bigl(1-2(mdl_3)^2\Bigl)K_1(2mdl_3)\Biggl]
\eea
and 
\bea
{\cal T}^{33 (11)}(d)&=& \lim_{p'\rightarrow p}\sum_{l_3=1}^\infty\frac{m \delta(p-p')(-1)^{l_3+l_0}\, e^{-4i(p_z-p'_z)dl_3}}{2\pi^2 \left(dl_3\right)^3}\Biggl[mdl_3(3-8imdl_3\gamma^3)\kappa_0(2mdl_3)\nonumber\\
&+&\Bigl(3+4mdl_3(mdl_3-2i\gamma^3)\Bigl)K_1(2mdl_3)\Biggl].
\eea
It is important to note that in this limit both the Casimir energy and pressure depend only on the distance $d$ between the plates. The dependence on Gamma matrices is not a problem since the formalism of the phase space is matricial. It leads to the conclusion that neither the energy nor the pressure are scalars but components of a tensor.

\section{Conclusion} \label{sec.4}

The Dirac field in phase space is considered. Using the Dirac equation, the propagator for spin-1/2 particles is calculated. This form of the propagator is similar to that in the usual quantum mechanics. The TFD results are obtained by using the temperature effects in the Dirac propagator. TFD, a real-time finite temperature formalism, is a thermal quantum field theory. Using this formalism, a physical (renormalized) energy-momentum tensor is defined. Then the Stefan-Boltzmann law in phase space and the Casimir effect are calculated at finite temperature. The results lead to the usual results for the Dirac field when they are projected in the quantum field theory space.
The TFD formalism allows studying the finite temperature effects in phase space. On the other hand such a formalism also may be used to explore the role of a thermal energy which is possibly related to the fermionic feature of the field.

\section*{Acknowledgments}

This work by A. F. S. is supported by CNPq projects 308611/2017-9 and 430194/2018-8.

\end{document}